\documentstyle[prl,aps]{revtex}
\twocolumn
\newcommand{\BEQ}{\begin{equation}}                                           
\newcommand{\EEQ}{\end{equation}}
\newcommand{\BEA}{\begin{eqnarray}}                              
\newcommand{\EEA}{\end{eqnarray}}

\renewcommand{\d}{{\rm d}}
\newcommand{\I}{{\cal I}}
\newcommand{\Ic}{{{\cal I}_c}}
\newcommand{\N}{{\cal N}}
\newcommand{\nn}{\nonumber\\}
\renewcommand{\S}{{S_{int}}} 

\begin{document}\draft
\title{Ehrenfest relations at the glass transition: 
solution to an old paradox} 
\author{  Th.  ~M.  ~Nieuwenhuizen }
\address{ Van der Waals-Zeeman Instituut, Universiteit van Amsterdam\\ 
          Valckenierstraat 65, 1018 XE Amsterdam, The Netherlands\\}
\date{July-5-1997; \today}
\maketitle
\begin{abstract}
In order to find out whether there exists a thermodynamic
description of the glass phase, the Ehrenfest relations along the glass 
transition line are reconsidered.
It is explained that the one involving the compressibility is always satisfied,
and that the one involving the specific heat is principally incorrect. 

Thermodynamical relations are presented for
non-ergodic systems with a one-level tree in phase space. 
They are derived for a spin glass model, checked for other models,
and  expected to apply, e.g., to glass forming liquids.
The second Ehrenfest relation gets a contribution from
the configurational entropy. 
\end{abstract}
\narrowtext

The glass transition is a dynamical freezing transition, that occurs when
a liquid is supercooled. The transition is smeared, but becomes the 
sharper the slower one cools. For ideal,  adiabatic cooling the 
transition will be sharp and occurs at the Kauzmann temperature $T_K$.

Experimentally it is known that second derivatives of the free 
energy, the specific heat, the 
compressibility and the thermal expansivity, make a (smeared) jump 
from their liquid values to smaller values in the glass. 
Since many decades in time are involved, one might
therefore  wonder whether it can be described 
as a (smeared) second order phase transition.
This idea has been put forward by Gibbs, DiMarzio and Adam.
As thermodynamics amounts to system-independent laws, 
the approach leads to sine-qua-non relations along the glass transition 
line $p(T)$. They are the Ehrenfest relations 
\BEA \label{Efrel1}
\Delta\kappa \frac{\d p}{\d T}&=&\Delta \alpha \\
\frac{\Delta C_p}{TV}&=&\Delta \alpha \frac{\d p}{\d T} \label{Efrel2}
\EEA
where $\Delta A\equiv A_{liquid}-A_{glass}$ for any $A$.
From experiments it was concluded that the first relation is 
usually violated, while the second is closely satisfied in 
most cases, but not in all.
(For reviews see \cite{Jackle},\cite{Angell}). 
The Prigogine-Defay ratio
\BEQ
\Pi\equiv\frac{\Delta C_p \Delta\kappa}{TV(\Delta\alpha)^2}
\EEQ
should be equal to $\Pi=1$. Experimental values are typically found, 
however, in the range $2 < \Pi < 5$. 
It is generally believed that $\Pi= 1$  is a strict lower bound. 

In an extension of the theory one assumes 
that at the transition a number of
(unspecified) internal variables $Z_i$ freeze in, 
and that the configurational entropy
is constant along the transition line.
This modifies (\ref{Efrel1}) and 
(\ref{Efrel2}) but keeps $\Pi=1$.~\cite{DiMarzio}
These negative results have prevented further development of a 
thermodynamic approach.

In this work we first explain that the first  Ehrenfest
relation is automatically satisfied (already in the reported 
measurements). By the same reasoning we shall
conclude that the second relation must be incorrect. Analyzing model systems 
we shall then derive an extra contribution that 
arises from the configurational entropy.

Before discussing the meaning of the Ehrenfest relations, we first have to
define the  experiment, or, better said, the set of experiments, 
to be performed.
Let us consider for definiteness cooling of  a glass forming liquid at a fixed 
pressure $p_1$ and cooling rate $Q=-\d T/\d t$. Starting
from a high temperature one measures the specific volume $V(T;p_1)$.
It is linear at large $T$ and at low $T$ and has a smooth crossover between
these behaviors. This happens  near the freezing  temperature $T_1$.
Let us then repeat the cooling experiment at a large set of different 
pressures $p_i$, with moderate steps $p_i-p_{i-1}$. 
This will lead to a set of freezing temperatures 
$T_i$, which define a smooth freezing line $p(T)$. 
The  location of this line is by no means universal.
It is defined by our set of measurements, here the set of $p_i$'s and their
common value $Q$ of the cooling rate.  
Different smooth sets of experiments may involve: a different cooling rate $Q'$; 
non-uniform cooling by letting $Q\to Q_i$ depend smoothly on $p_i$;
non-linear cooling; or cooling where also $p$ changes in time.  
All these sets of experiments will in principle lead to different
 transition  curves $p(T)$.
In practice this means that old works in literature, where the 
cooling procedure has not been specified, are not reproducible. 
Likewise, computer experiments, with their extremely high 
cooling rates, cannot 
 yield realistic glass transition temperatures.
 
To test the first Ehrenfest relation (\ref{Efrel1}) one needs  
$\kappa=-\partial \ln V/\partial p |_T$, which is difficult to
determine from cooling curves at two consecutive pressures.
Therefore it has become standard to measure $\kappa_1$ in the glass phase
 by cooling at $p_1$ down below
$T_1$ and then making small pressure variations. ~\cite{soundprop}
 Such procedures, however, lead to a determination
of $\d p/\d T$ from an experiment at (essentially) $p_1$ only,
 a contradictio inter terminis! 
{\it No experiment at one pressure can fix the slope of the transition line},
because that depends on the conditions under which the set of experiments will 
be performed. ~\cite{boilingline}
The hope that the compressibility could be obtained by small pressure
variations is  frustrated by the history dependence 
of the glassy state.  A closely related phenomenon is
known from experiments on spin glasses: the short-time (``zero-field cooled'')
susceptibility $\chi_{ZFC}=(1-q_{EA})/T$ is lower than the long-time 
(``field cooled'') susceptibility $\chi_{FC}$$=$$(1-\int_0^1\d xq(x))/T$, 
where $q_{EA}$ is the Edwards-Anderson order parameter and   
where $q(x)\le q_{EA}$ is the Parisi order parameter function.
In the glass the short-time value of $\kappa$ (measured by small pressure
steps~\cite{soundprop}) will also be too low, yielding
the observed too large $\Delta\kappa$.

The correct procedure is obvious
and comes from the meaning of the Ehrenfest relation.
The continuity of the specific volume can be considered at two glass 
transition points $(T_1,p_1)$ and $(T_2,p_2)$, where $\Delta V=0$.
One may thus write
$\Delta V(T_1,p_1)-\Delta V(T_1,p_2)$ $=$ 
$\Delta V(T_2,p_2)-\Delta V(T_1,p_2)$. 
(The terms at $(T_1,p_2)$ do not vanish). 
Dividing by $p_1-p_2$ and taking
the limit $p_2\to p_1$ this, of course, leads to eq. (\ref{Efrel1}). 
However, we must use on both sides of the equality 
the same values of $p_1-p_2$; it should not be taken ``infinitesimal'' 
on the left hand side and only ``small'' on the right hand side. 
In practice  {\it the only way} is to determine $\kappa$ from 
the curves $V(T;p_i)$. 
Modern computer graphics allows to fit all the
experimental data above and below the transition regions 
to  high- and low- $T$ surfaces in 
$V-p-T$ space. Using all data should lead to reasonable fits.
The intersection line of the surfaces will satisfy eq. (\ref{Efrel1}).
This approach thus explains the old paradox: If properly interpreted,
the first Ehrenfest relation is satisfied automatically! 
As there is no second procedure to produce the same glassy state,
there is hardly a point in testing (\ref{Efrel1}) experimentally. 

By the same token the second Ehrenfest relation (\ref{Efrel2}),
relating $dp/dT$ to measurements at one pressure only, 
cannot be correct! Likewise,  the Maxwell relation 
$\partial U/\partial p+p\partial V/\partial p=-T\partial V/\partial T $
 must be  violated in the glass.
We now show how unexpected behavior of the configurational entropy modifies
it. Below we discuss the derivation for spin glasses, and then extend it
to the glass transition in the hypernetted chain approximation.   
We believe that these relations are very general, 
and first formulate them for glass forming liquids.

For glassy systems the 
entropy consists of two terms. $\S$ is the internal entropy,
related to the glassy state the system condenses into; it lies 
well below the liquid entropy. $\Ic$ is the  configurational 
entropy due to the number of equivalent glassy states.
It is extensive at dynamical transitions, and becomes sub-extensive
only for ideal adiabatic cooling. 
This part of the entropy is ``lost'' 
in the glass transition region.
 The quantity $T\partial \S/\partial T|_p$ will generally 
be smaller than the specific heat $C_p=\partial (U+pV)/\partial T|_p$. 
In spin glass models the configurational entropy contributes
to the free energy as  $-(T/x)\Ic$, where $x$ is the weight 
$(0\le x \le 1)$ occurring
as breakpoint of the one-step replica symmetry breaking
 Parisi order parameter function.
$x$ also shows up dynamically as the factor by which 
the fluctuation-dissipation theorem is broken at long times.~\cite{CuKu}
$T_e\equiv T/x$ can be considered as effective temperature at which processes
related to the configurational entropy thermalize.~\cite{PelitiCuKu} 

It was shown that
$C_p=T\partial S_{int}/\partial T+T_e\partial\Ic/\partial T$.~\cite{Nthermo}
We are now in the position to take the first derivatives of the free
enthalpy  $G=U+pV-T\S-T_e\Ic$. This yields
\BEA\label{Gdp=}
\frac{\partial G}{\partial T}=-\S-\frac{\partial T_e}
{\partial T}\Ic; \qquad
\frac{\partial G}{\partial p}=V-\frac{\partial T_e}{\partial p} \Ic 
\EEA                                                                       
Along the transition line $G$, $S_{liq}=\S+\Ic$ and $V$ 
are continuous wrt to the liquid.
The standard assumption that the first derivatives of $G$ are also continuous,
is seen to be incorrect~\cite{Nmaxmin}: 
the terms involving $T_e$ are nontrivial in the glass 
($\partial T_e/\partial T<1$, $\partial T_e/\partial p\neq 0$).
 The finite difference in slopes discussed here is due to unexpected
 behavior of the configurational entropy,~\cite{Nmaxmin}
neglected so far. It leads to the modified Maxwell relation
\BEQ\label{dSdp}
\frac{1}{T}\frac{\partial U}{\partial p}
+\frac{p}{T}\frac{\partial V}{\partial p}
+\frac{\partial V}{\partial T}
=(\frac{T_e}{T}-\frac{\partial T_e}{\partial T})
\frac{\partial \Ic}{\partial p}
+\frac{\partial T_e}{\partial p}\frac{\partial \Ic}{\partial T}
\EEQ
Along the freezing line one has $T_e(T,p(T))=T$ and one
may define the total derivative $\d/\d T=\partial/\partial T+(\d p/\d T)
 \partial/\partial p$. Eq. (\ref{Gdp=}) does not violate the balance: 
$\d\Delta G/\d T=0$ since $\d T_e/\d T=1$. Let us now consider eq.
(\ref{dSdp}) and subtract  the values on the liquid-side.
Multiplying by $\d p/\d T$ and using $\d T_e/\d T=1$,
$\d\Delta U/\d T$$=$$\d \Delta V/\d T$$=0$ we  
 obtain from (\ref{dSdp}) the modification of the 
second Ehrenfest relation (c.f. eq. (\ref{Efrel2}))
\BEQ \label{modEf2}
\frac{\Delta C_p}{TV}
=\Delta\alpha\frac{\d p}{\d T}  +
(1-\frac{\partial T_e}{\partial T})\frac{\d \Ic}{V\d T}
\EEQ 
This relation indeed connects $\d p/\d T$ with another derivative 
along the transition line,
namely  that of the configurational entropy.
This term originates from the difference
in slopes of the liquid and glass free enthalpies.   
The factor $1-\partial T_e/\partial T>0$ is a nontrivial weight.
Since the first Ehrenfest relation is satisfied, measurement
of $\kappa$ is not needed for the Prigogine-Defay ratio:
$\Pi=\tilde \Pi\,\d T/\d p$ with $\tilde \Pi=\Delta C_p/( T V\Delta\alpha)$
determined at one pressure. $\Pi$ will
be less than unity when $\d\Ic/\d T<0$.

We now consider the data of Rehage and Oels 
for the glass transition of atactic polystyrene.~\cite{RehageOels}
 For cooling at a speed of $18\,K/hour$ at $p=1$ $bar$
 they report: $T=361$ 
$K$, $\Delta C_p/V=0.30$ $J/gK$, $\Delta\alpha=3.5\,10^{-4}$ $cm^3/gK$, 
$dp/dT=0.31$ $bar/K$, $\Delta\kappa=1.6\,10^{-5}$ $cm^3/g\,bar$.
This was reported to yield $\Pi=1.06\approx 1.0$, and a
violation of the first Ehrenfest relation. 
This violation has, however,
already been traced back to the way how $\kappa$ was measured. 
Our Prigogine-Defay ratio 
$\Pi=\tilde \Pi \d T/\d p =0.77$ {\it is less than unity}. The 
last term in eq. (\ref{modEf2}) is negative and brings  $23\%$ of the
value for the slope  $\d p/\d T$, a large effect.

We have discussed how the configurational entropy modifies an 
Ehrenfest relation. This effect should be stronger for first 
order glass transitions, which occur, for instance, in water.~\cite{Angell}
In the $p$-spin model (see eq. (\ref{Ham=}))
this happens when the transversal field $\Gamma$ exceeds a critical 
value.~\cite{NR} 
Using that along the dynamical transition line 
$\Delta G=\d\Delta G/\d T=0$ and eq. 
(\ref{Gdp=})  we obtain 
\BEQ
\Delta V\frac{\d p}{\d T}=\frac{1}{T}(\Delta U+p\Delta V)
+(\frac{T_e}{T}-\frac{\partial T_e}{\partial T})\Ic
\EEQ
which deviates by the $\Ic$ term from the static Clausius-Clapeyron equation.
Note that $x=T/T_e$ is below unity.
  
Let us now give the theoretical background of our relations.
They have initially been derived within a spherical $p$-spin interaction 
spin glass. For a system of $m$-component spherical spins $S_i^c$ 
($i=1,\cdots, N$, $c=1,\cdots, m$),
satisfying $\sum_{i,c} S_i^{c\,2}=N$,  
we consider the Hamiltonian in a transversal field
\begin{eqnarray}\label{Ham=}
{\cal H}=&-&\sum_{ i_1<\cdots<i_p} J_{i_1 i_2\cdots i_p}
S^z_{i_1}S^z_{i_2}\cdots S^z_{i_p}-\Gamma\sum_i S^x_i
\end{eqnarray}
The independent Gaussian quenched random couplings 
have average zero and variance $J^2p!/2N^{p-1}$. 
The system has a multitude of states $a=1,\cdots,\N$, each with its
own free energy $F_a$ that is a local minimum of a known
TAP-free energy functional.
The replica calculation with one-step replica symmetry breaking
involves parameters $\mu$, $q_d$,
$q_1$, and $x$, leading to the free energy ~\cite{NR}
\begin{eqnarray}\label{bFCS}
\frac{F}{N}&=&-\frac{\beta J^2}{4}(q_d^p-\xi q_1^p)
-\frac{T}{2x}\ln (q_d-\xi q_1) 
+\frac{T\xi}{2x}\ln(q_d-q_1)\nn
&+&\frac{\mu}{2}(q_d-1)-\frac{ \Gamma^2}{2\mu}+
\frac{(m-1)T}{2}\ln\frac{\mu}{T}
\end{eqnarray}
where $\xi=1-x$.
$\mu$, $q_d$, and $q_1$ are determined by optimizing $F$.
It was recently pointed out by us that the value of $x$ is related to
the time scale at which the system is considered.~\cite{Nthermo} Indeed,
setting  $\partial F/\partial x=0$ (leading to eq. (\ref{margeta}) with
$\eta\to \eta_{st}<1$ independent of $T$ and $\Gamma$)
yields the static phase transition at the
Kauzmann temperature $T_K$, related to the longest time scale.
On the other hand, the marginality condition 
(eq. (\ref{margeta}) with $\eta=1$)
describes algebraic time scales, at which a transition occurs at a higher
temperature $T_A$. This is reproduced by the  mode-coupling equations,
and is  comparable to the sharp critical temperature occurring in mode-coupling
equations for glasses. As this transition is absent in practice, 
we have considered the system at exponential time scales
$t=t_0\exp(N\tau)$.~\cite{Nthermo}
 At given $\tau$ barriers with free energy height less than
$NT\tau$ can be surpassed. 
$\eta$ parametrizes the value of the free energies of the states $a$.
As  time evolves, the dominant lowest reached free energy
$F_{min}(t)$ has parameter $\eta(t)$.
 When, at fixed field $\Gamma_1$, also the temperature $T(t)$ 
is slowly lowered, we can eliminate $t$ to obtain 
 a function $\eta(T;\Gamma_1)$.
Cooling trajectories at a large set of fields $\Gamma_i$ 
define a set of experiments. If the set is 
``smooth'' it will lead to a smooth function $\eta(T;\Gamma)$.
This function should follow from solving the dynamical equations. 
We shall not do that, but remain on a quasi-static level, where the
information of the cooling dynamics is coded in the function
 $\eta(T;\Gamma)$.

In our present analysis
a freezing transition occurs when the temperature, below  which
the dominant lowest reached state will freeze, 
is equal to the actual temperature. 
We assume that we can still describe the situation by the Gibbs weight, 
which is the case when no relevant parts of phase have become inaccessible.

The free energy of the TAP states can be  
characterized by a parameter $\eta$ ($\eta_{st}\le\eta\le 1$), 
that enters the condition 
\BEQ\label{margeta}
\frac{1}{2}\beta ^2 p(p-1) q_1^{p-2}
= \frac{\eta}{(q_d-q_1)^2}
\EEQ

In solving the saddle point equations for $\mu$, $q_d$, and $q_1$,
the above relation fixes $x=(p-1-\eta)(q_d-q_1)/\eta  q_1$.
One can calculate all quantities of interest.
We have verified the following relations for
$F(T,T_e(T,\Gamma),\Gamma)$: 
\BEA\label{S=S=}
F&=&U-T\S-T_e\Ic;
\qquad M=-\frac{\partial F}{\partial \Gamma}\vert_{T,T_e}\\
S_{int}&=&-\frac{\partial F}{\partial T}\vert_{T_e,\Gamma};\qquad 
\Ic=-\frac{\partial F}{\partial T_e}\vert_{T_e,\Gamma}\\
C&\equiv&\frac{\partial U}{\partial T}\vert_\Gamma=
T\frac{\partial S_{int}}{\partial T}\vert_\Gamma+
T_e\frac{\partial \I_c}{\partial T}\vert_\Gamma
\label{Ic=}
\EEA
This implies the modified Maxwell relation
\BEA
\frac{1}{T}\frac{\partial U}{\partial \Gamma}+\frac{M}{T}-\frac{\partial  M}
{\partial  T}=
(\frac{T_e}{T}-\frac{\partial T_e}{\partial T})
\frac{\partial \Ic}{\partial \Gamma}+
\frac{\partial T_e}{\partial \Gamma}\,
\frac{\partial\Ic}{\partial T}
\label{IdIp}
\EEA
$F$, $U$, $\S$, $\Ic$, $M$ and $T_e$ only depend on the
value of $\eta$ in the point $(T,\Gamma)$.
Their temperature derivatives
also  depend on $\partial \eta/\partial T$, while 
their field derivatives  
involve $\partial \eta/\partial p$, the measure of variation
between experiments at different fields.

The second Ehrenfest relation can be rederived by multiplying eq. 
(\ref{IdIp}) by $(1/N)\d \Gamma/\d T$. This yields generally
\BEQ \label{Ef2SG}
\frac{\Delta C}{NT}=\Delta\alpha \frac{\d \Gamma}{\d T} 
+ (1-\frac{\partial T_e}{\partial T})\frac{\d \Ic}{N\d T}
-(1-\frac{\d T_e}{\d T})\frac{\partial \Ic}{N\partial T}
\EEQ  
while $\Delta \alpha=\Delta\chi\,$$\d\Gamma/\d T$ is again
satisfied, where $\alpha\equiv -(\partial M/\partial T)/N$
and $\chi=(\partial M/\partial \Gamma)/N$.
In comparing with (\ref{modEf2}) one should keep in mind 
that $\d T_e/\d T=1$.
Eq (\ref{Ef2SG}) also applies to the unconventional
case where one starts cooling adiabatically below $T_K$. 
At freezing one then has 
$\Ic=\d\Ic/\d T=0$, but the last term is non-zero.

$\d\Ic/\d T$ does not depend on
$\partial \eta/\partial T$ but only on $\d \Gamma/d T$. 
It holds that $\Delta C$, $\Delta \alpha$ and $\Delta \chi$ are proportional 
 to $1-\partial T_e/\partial T$ 
(since near the transition $U_{sg}-U_{PM}\sim x-1$). 

The relations (\ref{S=S=}-\ref{Ef2SG}) are expected to be 
universal for dynamical glassy transitions  
with extensive configurational entropy. 
 We have considered three other cases:

1) In the case of a longitudinal field ($\Gamma\sum S_i^x\to H\sum S_i^z$)
the same equations are satisfied, see also~\cite{Nthermo}.

2) The hypernetted chain equation of fluids is an approximate non-linear 
integral equation for the pair-correlation function. ~\cite{vLGdB}
M\'ezard and Parisi~\cite{MezardParisi}
 pointed out that in a certain region it has many solutions, describing a
glass phase. By  weakly coupling different copies (replica's) of the
 system, they introduced a replica calculus. 
The main difference with the above spin glass is that the spin-spin
overlap is replaced by the pair  correlation function. 
 The static transition again follows from the relation 
$\partial F/\partial x=0$. 
Dynamically  this relation is not satisfied, and there is an
extensive configurational entropy. 
We now can make the same assumptions as in the above spin glass
and will rederive  eq. (\ref{modEf2}).

3) Recently we have introduced a model of a directed polymer on a
square lattice with a correlated random potential, 
consisting of randomly located
parallel ridges (repulsive potentials).~\cite{Ndirpol}
 The polymer prefers to lie in broad lanes (width $\ell$)
in between ridges, of which there occur a lot when the transversal width
scales as $W=\exp(\lambda L^{1/3})$ in the parallel width $L$.
The free energy reads
\BEQ\label{bFpol}
 F=Lf_B(T)+\frac{\Gamma(T)LT}{2\ell^2}-\nu\ell-T\Ic
\EEQ
where $f_B(T)$ is an uninteresting bulk free energy density, 
$\Gamma$ is the interface stiffness, $\exp(-\mu)$ is the chance for having 
no ridge at a given height, and $\nu$ is a chemical potential favoring
($\nu>0$) or disfavoring ($\nu<0$) wide lanes, and $\Ic=\log W-\mu\ell$
is the complexity. At some $T_K$,
where $\gamma(T;\nu)\equiv(\mu/\lambda)[T\Gamma(T)/(T\mu-\nu)]^{1/3}$ 
equals $\gamma(T;\nu)=1$, 
there is occurs a static ``Kauzmann'' transition from a 
high-temperature phase, 
where the polymer lies in the broadest lane
 ($\ell=\ell_{max}\equiv \lambda L^{1/3}/\mu$), to  a low
temperature phase, where it spends most of time in a set of narrower lanes  
($\ell^\ast=\gamma\ell_{max}$ with $\gamma<1$). 

Starting from a large set of uniformly distributed independent polymers,
we are interested in the dynamical (short time) regime A, where (\ref{bFpol})
is valid with $\ell$ increasing logarithmically with time.~\cite{Ndirpol} 
In order to make contact with previous theory, we 
replace $T\Ic$ in (\ref{bFpol}) by $T_e\Ic$ where $T_e(t)$$=$$
LT\Gamma(T)/(\mu\ell(t)^3)+\nu/\mu$.
 A minimum will then occur at $\ell=\ell(t)$.
In the polymer model one has a
reversed role of heating and cooling.~\cite{Ndirpol}
A set of experiments can be introduced by specifying smoothly related
heating trajectories at a large number of  $\nu_i$'s. For each of them
a dynamical phase transition can occur at any  $T<T_K(\nu_i)$, where
the dominant width $\ell(t)$ reached so far, equals its freezing value 
$\ell^\ast(T(t))$. This transition is very similar to the above ones.
Eqs. (\ref{S=S=}-\ref{Ef2SG}) are satisfied with $\Gamma\to \nu$.
As before, $\Ic$ is large at the transition.
The second term in the r.h.s. of eq. (\ref{Ef2SG}) is finite, while the 
last one vanishes. A related transition can occur at any $T>T_K$.
Then the transition line is $\ell(t)=\ell_{max}$, where
$\Ic=0$. Now the last term in (\ref{Ef2SG}) is non-vanishing.

In conclusion, we have pointed out that the present understanding
of the Ehrenfest relations is incorrect. We have explained that the first
one is satisfied automatically and that the second one must be modified.
 From quasi-static model calculations we have shown that it
 gets an extra contribution from the configurational entropy.
This explains that the Prigogine-Defay ratio is smaller than unity
in the experiments or ref. ~\cite{RehageOels}

We have also presented the generalization of the standard 
thermodynamical laws
to non-ergodic situations with a one-level tree in phase space.
 This is given in eqs. (\ref{S=S=})-(\ref{IdIp}), where
$T_e$ is an effective temperature, that depends slowly on time.
Along the transition line the modified Maxwell relation
eq. (\ref{IdIp}) leads to a new form
(\ref{modEf2},\ref{Ef2SG})  of the second Ehrenfest relation. 
We have checked the predictions in several model systems.

Note added: we have checked  our Ehrenfest relations for cooling
in the backgammon model, which has no disorder but
entropic barriers.~\cite{Ritort}

\acknowledgments 
The author thanks
S. Franz,  
D. Frenkel,  
H.F.M.  Knops, 
W.A. van Leeuwen, 
J. Mi\-chels,  
B. Nien\-huis, 
K.O.  Prins, 
F. Ri\-tort, 
Th.W. Ruijg\-rok, 
J.A.  Schouten, 
and  
G.H.  Wegdam 
for discussion, and the ISI (Turin, Italy)
 for hospitality.

\end{document}